\PassOptionsToPackage{table}{xcolor}
\documentclass[sigconf]{acmart}
\usepackage{colortbl} 
\usepackage[capitalise]{cleveref}
\usepackage{tcolorbox}

\newcommand{\interviewquote}[2]{
 \def\FrameCommand{%
    \hspace{0pt}%
    {\color{cyan} \vrule width 2pt}
    \colorbox{white}
  }%
  \MakeFramed{\advance\hsize-\width\FrameRestore}%
  \noindent
  \begin{adjustwidth}{}{1pt}
  {\small``\textit{#1}'' - {#2}}\end{adjustwidth}\endMakeFramed%
}

\AtBeginDocument{%
  }

\copyrightyear{2026}
\acmYear{2026}
\setcopyright{cc}
\setcctype{by-nc-nd}
\acmConference[ICSE-SEET '26]{2026 IEEE/ACM 48th International Conference on Software Engineering}{April 12--18, 2026}{Rio de Janeiro, Brazil}
\acmBooktitle{2026 IEEE/ACM 48th International Conference on Software Engineering (ICSE-SEET '26), April 12--18, 2026, Rio de Janeiro, Brazil}
\acmPrice{}
\acmDOI{10.1145/3786580.3786982}
\acmISBN{979-8-4007-2423-7/2026/04}

\begin{document}

\title[Design Thinking on Gender Equity]{From Personas to Programming: Gender Preferences in Design Thinking-Based Computing Education at Secondary Schools}
\title[Design Thinking on Gender Equity]{From Personas to Programming: Gender-specific Effects of Design Thinking-Based Computing Education at Secondary Schools}

\author{Isabella Gra{\ss}l}
\email{isabella.grassl@tu-darmstadt.de}
\orcid{0000-0001-5522-7737}
\affiliation{%
  \institution{Technical University of Darmstadt}
  \city{Darmstadt}
  \country{Germany}}

\author{Gordon Fraser}
\email{gordon.fraser@uni-passau.de}
\affiliation{%
  \institution{University of Passau}
  \city{Passau}
  \country{Germany}}

  \author{Daniela Damian}
\email{danielad@uvic.ca}
\affiliation{%
  \institution{University of Victoria}
  \city{Victoria}
  \country{Canada}}

\begin{abstract}
Creative approaches to attract students to software engineering at an early age are emerging, yet their differential impact on gender remains unclear. 
This study investigates whether design thinking's empathy-driven approach addresses the documented gender gap in interest in software engineering. 
In a 10-week curriculum-integrated design thinking software development course with 55 secondary school students aged 13-–15 from two schools in Canada, we examined gendered differences in perceived gains in knowledge and interest, as well as in social-emotional experiences.
Our results show that both girls and boys gained perceived knowledge in software development. However, girls showed significant improvements in self-efficacy, interest, engagement with sustainability topics, and well-being, including optimism, sense of usefulness, and social connectedness. Positive emotions were strongest during creative, collaborative phases, while technical tasks led to some boredom, especially among boys, though they still benefited overall. 
This suggests that human-centred design thinking might be one effective way to address gender equity challenges, though we need more differentiated technical implementations.
\end{abstract}



\maketitle

\section{Introduction}
\label{sec:intro}
Gender gaps in software engineering education persist despite efforts to increase participation and equal technical abilities between girls and boys, particularly during early adolescence~\cite{sun2022programming,master2021d,spieler2020female}. 
These gaps often reflect misalignments between what traditional curricula and what learners, particularly girls, find meaningful and engaging~\cite{peixoto2018}. 
For many girls, software engineering can feel abstract, impersonal, and disconnected from real-world impact, their values, and social concerns~\cite{master2021d,spieler2020female}.

To address these limitations, researchers have advocated for more holistic approaches that integrate creativity, collaboration, and societal relevance \cite{fidjeland2023a,morales-navarro2022,lisi2024}. 
One promising method is \emph{design thinking}: a human-centred, iterative problem-solving process rooted in empathy, creativity, and collaboration~\cite{stanfordd.school.2017}. 
Rather than replacing technical instruction, design thinking frames programming and system design within the whole software engineering cycle, from understanding users to ideating, prototyping, and testing. 

Design thinking has shown promise in promoting student agency and engagement across disciplines~\cite{possaghi2024a,chen2017}. 
Notably, the emphasis on empathy and collaboration aligns with research showing that girls tend to be more attracted to creative subjects that align more with their life experiences~\cite{cheryan2015,spieler2020female}. This suggests that design thinking may be especially effective in engaging girls by valuing perspectives and competencies that are often under-represented in traditional computing environments~\cite{margolis2002}.


While previous research has explored gender-specific effects of approaches such as block-based programming and creative computing for engagement~\cite{fidjeland2023a}, focusing on girls~\cite{jamshidi2024,grassl2024girls}, few studies have examined how girls and boys  experience structured, design thinking-based software development processes~\cite{aguilera2021a}, especially when addressing purposeful topics such as sustainability and socially relevant issues to particularly motivate girls by connecting those themes with technical skills~\cite{peters2024,fields2023,nguyen-duc2023,spieler2020female,aguilera2021a}. 
Given the human-centred nature of design thinking, we would expect that such approaches benefit girls’ engagement in particular~\cite {blikstein2019a}; however, there is limited empirical evidence to date.

To address this gap, we conducted a 10-week design thinking-based software development course using MIT App Inventor~\cite{patton2019}, following best practices~\cite{stanfordd.school.2017,grover2013a}, with 55 secondary students aged 13--16 from two Canadian schools.
To understand how this approach might support broader participation in software development, we investigate four interconnected outcomes:  (1) Perceived knowledge gains~\cite{alsmadi2024} provide the foundation for developing self-efficacy~\cite{bandura1977a}, which in turn fosters (2) interest~\cite{bussey2004a}; (3) emotional experiences during different phases of design thinking, which shape motivation and persistence~\cite{zhang2022b}; and (4) students’ overall perceived well-being~\cite{anthony2022}, capturing the broader socio-emotional impact of engaging with meaningful, collaborative, and creative software projects. This allows us to investigate the following research questions:

\textbf{RQ1:} How do girls and boys perceive their knowledge and self-efficacy before and after the course?

\textbf{RQ2:} How do girls and boys perceive their interest in sustainability and software development before and after the course?

\textbf{RQ3:} What emotions do girls and boys associate with different phases of the design thinking process?

\textbf{RQ4:} How do girls and boys perceive their well-being before and after the course?

While students of all genders benefited from the course, girls showed statistically significant gains across all measured outcomes, including perceived knowledge, self-efficacy, interest, and well-being. Boys also improved, particularly in knowledge and self-efficacy, though their gains were less pronounced. 
Notably, introducing software engineering through design thinking enabled students to engage meaningfully with sustainability topics, fostering agency and positioning them as active problem solvers.

This work provides the first empirical evidence comparing gender differences in design thinking-based software engineering education, demonstrating that such approaches may be  practical for supporting girls’ engagement in software engineering. 

\section{Background and Related Work}

\subsection{Gender in Software Engineering Education}
\textbf{Gender disparities} in software engineering education remain a pressing issue~\cite{lewis2016don,weeden2020pipeline,sun2022programming}. Despite equal or stronger academic performance, girls often report lower self-efficacy, reduced interest, and a limited sense of belonging~\cite{spieler2020female}. Traditional computing curricula typically focus on abstract, technical skills, which may not resonate with students seeking creativity or social relevance~\cite{peixoto2018,fields2023,wit2023,scott2023a}. 

In contrast, design-based and personally meaningful approaches have been found to increase engagement, particularly among girls~\cite{fidjeland2023a,aguilera2021a,Jawad2018,Terroso2022}, resulting in diverse frameworks for gender-sensitive activities such as \emph{PECC} (Playing, Engagement, Creativity, Creating)~\cite{spieler2025}.
Girls, especially in middle school, are more likely to participate in programming activities when they see its relevance to their lives and values, and when learning environments foster emotional safety, creativity, and shared purpose~\cite{master2021d,spieler2020female,denner2011}. 

\textbf{Cultural context} also shapes the gender gap in software engineering education. While disparities are pronounced in the US, studies with middle and high school students  in countries such as India and Saudi Arabia show minimal differences in girls’ participation and self-efficacy compared to boys~\cite{chen2024exploring}. Similarly, research with adolescents aged 13 to 15  in Slovenia and the Middle East demonstrates that, despite both groups enjoying gamified, socially relevant learning programming, participants from the Middle East achieved higher post-game scores. At the same time, Slovenians appreciate collaborative work more~\cite{kramar2025}. However, e.g., White and Latina girls in California are most encouraged to pursue computing classes and careers when they perceive clear \emph{value} in computing during middle school~\cite{denner2011}.

\subsection{Relevance of Affective Factors}
Affective experiences are essential in young learners’ engagement and motivation in programming activities, particularly for girls~\cite{tsan2023}.

\textbf{Perceived Knowledge and Self-Efficacy.}  
Students’ perception of their own knowledge often precedes and shapes self-efficacy~\cite{tisza2023,morales-navarro2023a,clarke-midura2019b}. Self-efficacy~\cite{bandura1977a}, defined as a student’s belief in their ability to succeed, is strongly influenced by emotional experiences and prior knowledge~\cite{clarke-midura2020a,pekrun2019}. Positive experiences with problem-solving, collaboration, and meaningful tasks can increase self-efficacy, while repeated frustration or failure may reduce confidence. In computing education, self-efficacy predicts persistence, engagement, and future participation, and is particularly relevant for underrepresented groups such as girls~\cite{spieler2020female,master2021d,aguilera2021a}.

\textbf{Interest.} 
Learning often progresses from knowledge acquisition to interest development, which then supports self-efficacy~\cite{tisza2023,morales-navarro2023a,clarke-midura2019b}. According to Expectancy-Value Theory, motivation depends not only on expectations of success but also on the value students assign to a task~\cite{eccles1983,wigfield2000}. Situated learning frameworks emphasise that foundational knowledge enables learners to form meaningful connections with tasks, fostering curiosity, interest, and confidence~\cite{lave1991,eccles2020,clarke-midura2019b,tellhed2022b}. In programming education, where tasks can be abstract, fostering interest is critical to supporting self-efficacy~\cite{tisza2023,theodoropoulos2022b}.

\textbf{Emotions.}  
Novice programmers often experience anxiety, particularly in text-based programming environments compared to block-based ones~\cite{yusuf2024}. Emotions influence how students interpret their performance, shaping motivation and engagement~\cite{pekrun2019,morales-navarro2023a}. Positive emotional experiences are significant for girls, linking to sustained interest and confidence~\cite{tsan2023,spieler2020female,master2021d,aguilera2021a}. Approaches such as design thinking, which prioritise empathy, creativity, and relevance, can foster the emotional engagement necessary for long-term motivation~\cite{clarke-midura2020a}.

\textbf{Well-being.}  
Well-being is an essential but under-studied factor in software engineering (education)~\cite{takaoka2024}. Key components include \emph{environmental mastery, personal growth, purpose, autonomy, self-acceptance, and positive relationships}~\cite{takaoka2024}. Students at the university level can experience stress from debugging challenges, problem-solving failures, and feelings of imposter syndrome~\cite{rosenstein2020,chandrasekaran2025}. Supporting mental health and well-being in software development activities can improve engagement, confidence, and sustained interest~\cite{wong2023mental,montes2025wellbeing,takaoka2024}.

\subsection{Design Thinking Approach}
Traditional software engineering education often prioritises individual technical skill acquisition, with limited creative freedom and little connection to real-world contexts, due to instructor-defined tasks~\cite{tedre2018,blikstein2019a}. In contrast, socially- and human-centred projects support engagement by allowing students to see the positive impact of their work and having creative freedom~\cite{williams2024,alsmadi2024}. 

Design thinking offers a collaborative, creative framework built around empathy, ideation, and iterative problem-solving~\cite{stanfordd.school.2017}. Its five phases (empathise, define, ideate, prototype, and test) emphasise real-world problems and user perspectives, aligning well with inclusive computing education goals~\cite{stanfordd.school.2017,clarke-midura2020a}. 

Research in K-12 settings shows that design thinking can enhance engagement, foster solution-oriented thinking, and reduce gender disparities in outcomes~\cite{possaghi2025,yen2021,zhang2022b}. However, most of these activities focused on physical game-building~\cite{zhang2022b}, hands-on engineering tasks~\cite{yen2021}, or cybersecurity exercises~\cite{possaghi2025}, without involving actual programming, and therefore did not provide students with experience of the whole software development process.

\subsection{Block-based Programming}
Block-based programming tools play a complementary role by lowering technical barriers and allowing beginners to focus on problem-solving and computational thinking~\cite{nikou2014,papadakis2017}. MIT App Inventor enables students to create mobile apps through drag-and-drop programming~\cite{patton2019}. Its visual interface, combined with a project-based, real-world focus, has proven effective in engaging and motivating students and in learning fundamental computing concepts~\cite{nikou2014,grover2013a,tissenbaum2018,papadakis2019a}. 

App Inventor projects often explore socially relevant or community-based themes, which can be especially motivating for girls and other underrepresented groups~\cite{clarke-midura2020a,peixoto2018,motschnig2017,peters2024}. Sustainability-oriented computing projects tap into students’ social and ecological concerns, helping them see programming as a tool for positive impact~\cite{motschnig2017,wong2023}. Such themes align naturally with design thinking’s emphasis on empathy and human-centred design. Integrating design thinking with block-based programming and meaningful themes creates a promising pathway for inclusive software engineering education.


\section{Method}

In order to answer the four central research questions, we conducted a 10-week course with 55 students aged 13--16 from two secondary schools in Canada, focusing on sustainability-oriented app design using design thinking and block-based programming. 
Our analysis provides insight into how design thinking shapes students’ perceptions of software and sustainability (\emph{RQ1--RQ2}) as well as their socio-emotional development (\emph{RQ3--RQ4}).

\subsection{Course Design}
\label{sec:course}
The course is grounded in constructionist learning theory \cite{papert2020}, promoting learning through making, emphasising problem-solving and creativity~\cite{spieler2020female}.
Our intervention aims at encouraging students to creatively explore the software engineering process 
through design thinking~\cite{stanfordd.school.2017}, positioning them as active problem-solvers addressing sustainability-related challenges~\cite{penzenstadler2018,motschnig2017}, rather than focusing solely on technical skills.

To make the course gender-sensitive, we drew on the \emph{PECC} framework~\cite{spieler2025}. \emph{Playing} is interpreted broadly, giving students freedom to choose what to implement; we introduce possibilities by showing example projects from the App Inventor website. \emph{Engagement} is supported through collaboration, communication, and tinkering, allowing students to work together and experiment actively. \emph{Creativity} is encouraged by combining structured design tasks with the freedom to personalise projects, fostering self-expression and incorporating artwork or design-focused elements. \emph{Creating} involves hands-on programming, allowing students to experience the whole software development process while completing interdisciplinary projects.
By incorporating these elements, the course aims to enhance perceived knowledge gain, self-efficacy, interest, and enjoyment~\cite{spieler2025}, while also supporting well-being.

The course spans ten weeks, providing sufficient time for secondary school students to learn both design thinking and app development at an age-appropriate pace. The structure allows step-by-step guidance through the complete design thinking process, while remaining flexible enough to accommodate more advanced app features if additional time is available.

\subsection{Course Context}
The course was conducted in secondary school classrooms in Canada, with one class each from a public and a private school. Although situated in formal education, the design is transferable to other contexts, including extracurricular activities. Embedding the intervention in formal education was a deliberate choice to ensure inclusivity and to observe how students engage with novel topics in their everyday learning environment. None of the participants had received any formal computer science courses before the study.

The researchers initiated collaboration with schools, presenting the study goals and course content to teachers. The teachers provided feedback to ensure alignment with the school context. In practice, teachers were responsible for forming the teams and were present in most sessions but primarily served a supportive role, answering questions, supervising, and offering feedback, while the research team facilitated the course. Their involvement complements the intervention, as they know the students best, yet the design remains feasible without heavy teacher participation, making it adaptable across different settings.  

\subsection{Course Structure}
The course is structured to follow the five phases of the design thinking process~\cite{stanfordd.school.2017}, with integrated activities that build technical, collaborative, and reflective skills over time~\cite{chatzinikolakis2014,grover2013a,tissenbaum2018}.

Each session lasts approximately two to three hours and focuses on a distinct phase, with the intention of developing students’ awareness of sustainability issues as a motivating context that connects technical skills to societal impact~\cite{tedre2018}, strengthening their self-efficacy in computing, and fostering positive emotional engagement with the software development process~\cite{tsan2023,grassl2022gender}.

We provide all materials including the activities, lesson plans, and the analysis including codebook in our replication package.\footnote{\url{https://figshare.com/s/fd0324bf0b5cc660e776}}

\begin{figure}
	\centering
	\includegraphics[width=1\linewidth]{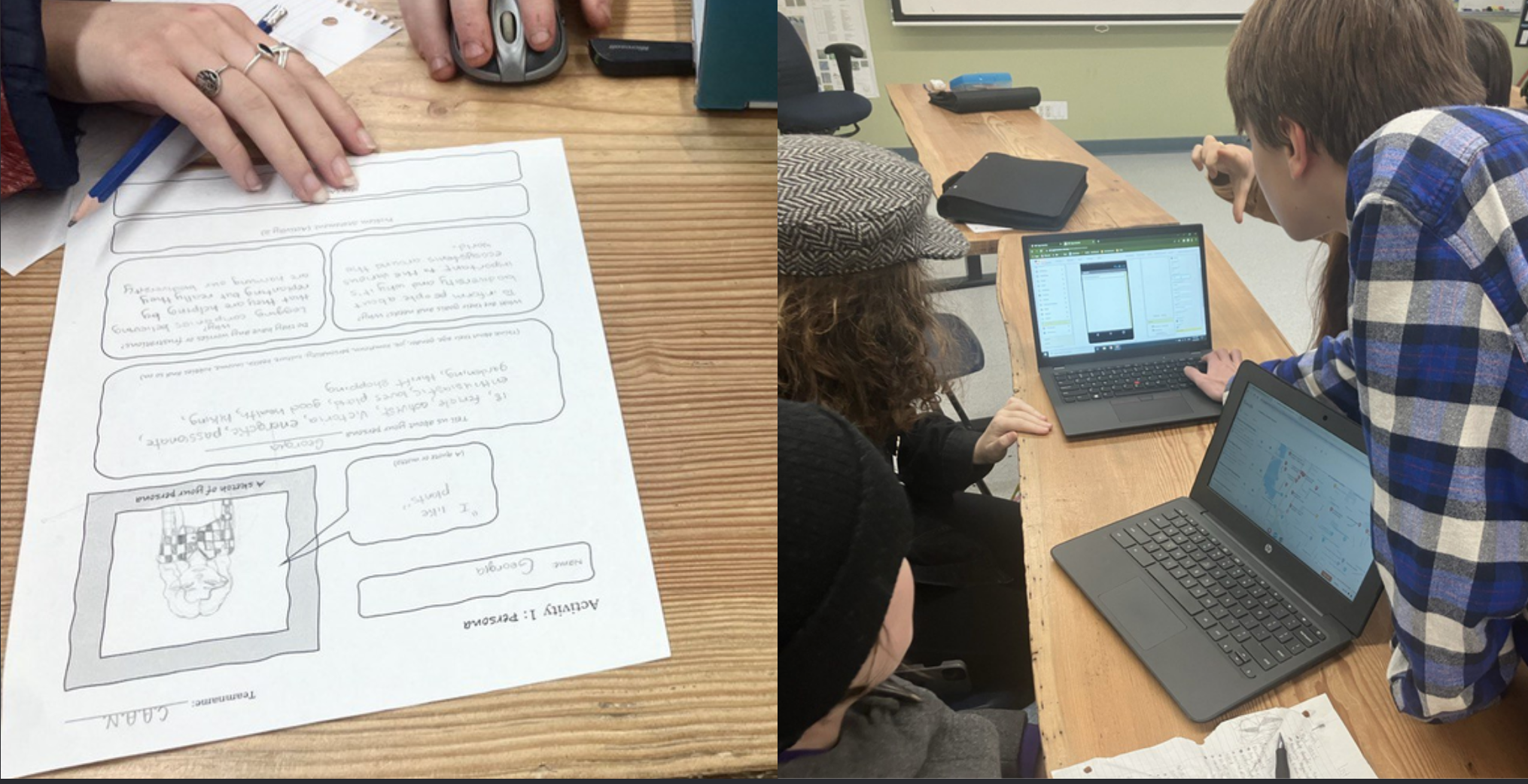}
	\caption{Student team developing their persona (session 4, left) and designing their app interface (session 8).}
	\label{fig:example}
\end{figure}

\subsubsection{Introduction: Design Thinking \& Software Development (Session 1--2)}
Students are introduced to design thinking and software development, explaining that software development can be seen not merely as technical problem-solving, but as a creative and collaborative process grounded in human needs. We also include icebreakers and team-building activities.


\subsubsection{Empathize Phase (Sessions 3--4)}
The \emph{brainstorming} session (session 3) focuses on identifying sustainability-related issues that students care about. Students are tasked to think about their everyday lives, communities, and global concerns. This activity is essential in ensuring that the subsequent development process is personally meaningful and socially grounded. 
In the \emph{Persona Development} session (4), students develop personas to foster empathy and user-centred thinking. They conduct short, informal interviews with classmates from other classes and teachers. They fill out the first activity sheet, which includes demographic information, goals, challenges, and needs, for each persona (\cref{fig:example}). This activity helps students understand that technology is designed for people with diverse needs and experiences.

\subsubsection{Define Phase (Session 5)}
Students use insights from the persona work to define specific problems they want to solve. They synthesize their research into point-of-view statements and create \emph{problem frames} that include the user, their need, and the insight driving the design on the second activity sheet. This stage emphasize analytical thinking and helps students narrow down their broad ideas into feasible design goals.

\subsubsection{Ideate Phase (Session 6)}
The ideation phase challenges students to develop a wide range of possible solutions using techniques such as mind mapping, card sorting, and sketching on their third activity sheet. Students are encouraged to defer judgment, build on others’ ideas, and think divergently. 
Ideas are then clustered and voted on by the team, culminating in the selection of one or two core features to pursue. This phase promotes creativity, ownership, and a deeper engagement with the problem space.
The resulting ideas covered a wide range of socially relevant topics, such as: 
\begin{itemize}
    \item Managing volunteering for students to clean up the ocean, possibly targeting those who need volunteering hours.
    \item Simplifying volunteering accessibility for busy young people, with multilingual support and centralized information.
    \item Helping women and kids under 16 feel safer in public.
    \item Educating children in hospitals about health.
    \item  A recycling game to teach kids about recycling.
    \item An app for teachers and the public to learn and donate to promote biodiversity.  
\end{itemize}


\begin{figure}
	\centering
	\includegraphics[width=1\linewidth]{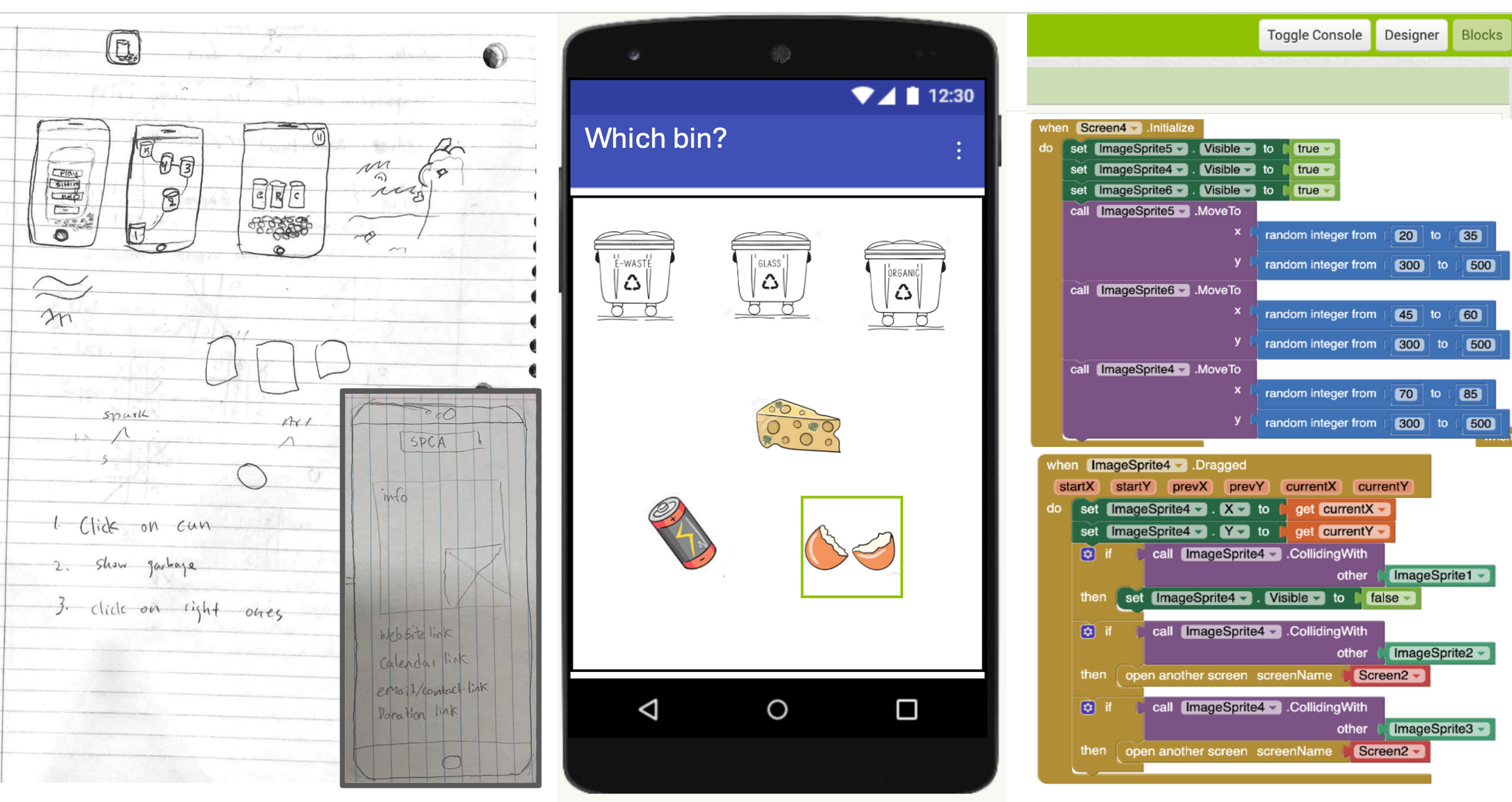}
	\caption{Recycling sorting game: scratch to final app.}
	\label{fig:exampleFINAL}
    \vspace{2em}
\end{figure}

\subsubsection{Prototype \& Test Phase (Sessions 7--9)}
\paragraph{Low-fidelity Prototyping (Session 7)}  
In the prototyping session, students develop paper prototypes of their app interfaces and act out user scenarios through role-playing and \emph{think-aloud} testing. The low-fidelity prototypes allow for rapid iteration and critique, with feedback loops built into the session. Students reflect on how different designs influence user perception.

\paragraph{Interface Design (Session 8)}  
Students begin implementing their designs using MIT App Inventor (\cref{fig:example}). The focus is on layout, navigation, and user interface elements. Students receive targeted support on visual design principles. The design phase aims to reinforce the connection between user needs and technical realisation, while being fully creative and expressing oneself~\cite{grover2015}.

\paragraph{Programming (Session 9)}  
This session continues the development phase, now focusing on the apps' logic and functionality. Using the block-based interface in App Inventor, students implement basic features such as buttons and conditional logic. This technical work is framed not as an isolated challenge but as part of the iterative design process. Many students return to earlier stages to revise their ideas based on implementation feasibility.

\paragraph{Presentation and Reflection (Session 10)}  
Students present their apps, reflecting on their process and design decisions, and sharing their learning journeys. The session celebrates progress and provides a platform for peer learning and feedback.

\subsection{Participants}
A total of 55 students ($31$ girls, $24$ boys, $0$ non-binary) took part in the course. The majority were aged 14 ($n = 32$), with smaller groups aged 13 ($n = 10$), 15 ($n = 5$), and 16 ($n = 8)$. 
Participants had diverse ethnic backgrounds, including European ($28$; 14 girls), Asian ($10$; 8 girls), North American ($11$; 4 girls), African ($2$; both girls), South American ($1$; girl), Middle Eastern ($2$; both girls), and one boy identifying as both European and First Nations.

\subsection{Instrumentation}
\begin{table}[tb]
\centering
\small
\caption{Questionnaire for the students.}
\vspace{-1em}
\label{tab:designthinkgindemos}
\begin{tabular}{@{}p{0.5cm}p{5.4cm}p{0.8cm}p{0.9cm}@{}}
\toprule
Var. & Question & Type & Source \\
\midrule
\multicolumn{4}{l}{Perceived Knowledge and Self-Efficacy (RQ1)}\\
\midrule
    KN01 & I have programmed before. & 5-pt-LI & \cite{grassl2023g}\\ 
    KN02 & I have developed a software project before. & 5-pt-LI& \cite{penzenstadler2018} \\ 
    KN03 & I know the general software development process (the phases, roles). & 5-pt-LI&\cite{penzenstadler2018} \\ 
    KN04 & I know what design thinking is about. & 5-pt-LI&\cite{penzenstadler2018}\\ 
\midrule
    SE01 & I think that I can solve problems in developing software by myself. & 5-pt-LI& \cite{agarwal2000,schwarzer1995}\\ 
\midrule
\multicolumn{4}{l}{Perceived Interest (RQ2)}\\
\midrule
    IN01 & I am interested in software development. & 5-pt-LI & \cite{munoz2016a}\\ 
    IN02 & I would consider developing software as (part of) my profession. & 5-pt-LI& \cite{spieler2020female,munoz2016a}\\ 
    IN03 & I am interested in topics related to sustainability. & 5-pt-LI & \cite{penzenstadler2018}\\ 
    IN04 & I am participating in projects related to sustainability. & 5-pt-LI & [new]\\ 
    IN05 & I feel empowered to actively contribute to sustainability topics. & 5-pt-LI &  \cite{penzenstadler2018,pekrun2011}\\ 
    IN06 & How do you think software engineering could help (if at all) achieving sustainability? & FT & [new] \\
\midrule
\multicolumn{4}{l}{Emotional Engagement (RQ3)}\\
\midrule
    EM01 &What emotions do you have when you think back on the course experience for each phase? & 8 Emo.& \cite{pekrun2011}\\ 
    EM02& Please briefly explain your choices. &FT&[new]\\ 
\midrule
\multicolumn{4}{l}{Perceived Well-Being (RQ4)}\\
\midrule
    WB01 & I've been feeling optimistic about the future. & 5-pt-LI &  \cite{tennant2007,anthony2022}\\ 
    WB02 & I've been feeling useful. & 5-pt-LI &  \cite{tennant2007,anthony2022}\\
    WB03 & I've been feeling relaxed. & 5-pt-LI &  \cite{tennant2007,anthony2022}\\ 
    WB04 & I've been dealing with problems well. & 5-pt-LI &  \cite{tennant2007,anthony2022}\\ 
    WB05 & I've been thinking clearly. & 5-pt-LI &  \cite{tennant2007,anthony2022}\\ 
    WB06 & I've been feeling close to other people. & 5-pt-LI & \cite{tennant2007,anthony2022}\\ 
\bottomrule
\end{tabular}
\end{table}

In this study, we focus solely on students’ self-reported experiences rather than measuring \emph{actual} learning outcomes~\cite{zhang2022b}. While self-reports are inherently subjective and may not perfectly reflect reality, our goal is to capture students’ perspectives and give them a voice. This approach also encourages engagement with the survey, motivating the students~\cite{robins2007}. Although the coherence between self-reported and “objective” learning experiences is not always evident, research has shown that students’ perceptions are often related to their learning outcomes and behaviours~\cite{han2022, pardo2016,zhang2022b}. 
To ensure the validity of our self-report items, we followed established guidelines for questionnaire design~\cite{kuh2021}, creating questions that students could answer, that were clear, non-threatening, meaningful, and focused on recent experiences. 

\Cref{tab:designthinkgindemos} shows the questionnaire, where the majority of items were adapted from prior studies to ensure comparability and relevance to our course context.
We used single-item measures to reduce participant burden and keep the questionnaire concise. Prior research has shown that specific constructs (e.g., satisfaction, self-esteem, flourishing) can be reliably assessed with single items~\cite{cheung2014,robins2001,burns2025}.

\paragraph{RQ1: Perceived Knowledge \& Self-Efficacy} 
Students’ prior experience and knowledge shape how they approach (un)familiar, challenging tasks~\cite{zhang2022b,ko2025}. Contextualising their perceived knowledge allows instructors to understand which aspects of software engineering and human-centred design are novel or familiar~\cite{penzenstadler2018,grassl2023g}, and how this might influence motivation, persistence, and allow for effective scaffolding~\cite {shamir2019, jamshidi2024}. 

Self-efficacy reflects students’ confidence in solving problems, which influences persistence when facing challenging tasks and willingness to experiment creatively, which are key aspects of both software development and design thinking~\cite{agarwal2000, schwarzer1995}. Therefore, we measured prior programming and project experience (\emph{KN01}, \emph{KN02}), knowledge of software development (\emph{KN03}) and design thinking (\emph{KN04}), and self-efficacy (\emph{SE01}, \cref{tab:designthinkgindemos}).

\paragraph{RQ2: Perceived Interest}
Interest drives engagement, learning, and career aspirations, particularly when connecting stereotypical technical work to socially relevant outcomes~\cite{spieler2020female}. We argue that fostering interest in software engineering and sustainability encourages students to see beyond technical tasks and engage in human-centred problem-solving~\cite{pekrun2011, penzenstadler2018}. 
We assessed interest in computing (\emph{IN01}), career aspirations (\emph{IN02}), sustainability-related projects (\emph{IN03}--\emph{IN04}), sense of empowerment (\emph{IN05}), and reasoning about why software engineering and sustainability fit together (\emph{IN06}). This helps capture motivation and values that support sustained engagement. 

\paragraph{Emotional Engagement (RQ3)} 
Emotions strongly influence learning, creativity, and collaboration, which are also central to design thinking~\cite{takaoka2024,zhang2022b}. Positive emotions like hope and pride can enhance engagement and problem-solving, while negative emotions such as anxiety or frustration highlight where students struggle~\cite{chandrasekaran2025,yusuf2024}. Measuring emotions across the project phases allows us to link affective experiences to design thinking activities, empathy development, and interest in software development~\cite{pekrun2011}. Students rated eight emotions: happiness, hope, pride, boredom, anger, anxiety, sadness, and fear (\emph{EM01}), and explained their ratings in an open-ended response (\emph{EM02}).

\paragraph{Perceived Well-Being (RQ4)} 
Well-being supports students’ ability to engage, reflect, and persist in projects, yet it is rarely studied in software engineering education despite its importance. In projects, students must navigate the uncertainty of user requirements, discussing social and technical issues, and collaboration, making well-being relevant for motivation and resilience. We measured six aspects of mental well-being, optimism, sense of usefulness, relaxation, problem-solving confidence, clarity of thought, and social connectedness, using a validated short scale~\cite{anthony2022} that is based on the ~\emph{Warwick–Edinburgh Mental Well-being Scale}~\cite{tennant2007} (\emph{WB01}--\emph{WB06}).

\subsection{Data Collection}
Data were collected using pre- and post-course questionnaires and open-ended reflections (\cref{tab:designthinkgindemos}). The pre-course collection took place right before the first session started, except for \emph{IN06} and \emph{EM01--02}. Since we assumed that many students do not know what software engineering encompasses (\emph{IN06}), we provided a brief explanation that this means designing, implementing, and testing software applications, then handed out the question for \emph{IN06}. We assessed the emotions right after each phase (\emph{EM01--02}). The post-course collection took place after the presentations of the last session.
Surveys were pseudonymized via student-selected nicknames. Demographic data (age, gender, ethnicity) were collected, and informed consent was obtained from parents, teachers, and schools. Students were briefed on the study’s goals and ethical principles.

\subsection{Data Analysis}
To compare responses across girls and boys, we use the \emph{Mann-Whitney U test}, a non-parametric test suitable for small, independent samples. 
For within-group analysis (pre–post), we use the \emph{Wilcoxon Signed-Rank test}, which is appropriate for paired ordinal data, at $\alpha \leq 0.05$. 
For both comparisons (\emph{girls vs boys}, \emph{pre vs post}), we report Vargha-Delaney’s $\hat{A}_{12}$ effect sizes. Values above $0.5$ indicate that the first group tends to score higher, and values below $0.5$ suggest that the second group tends to score higher. 
In gender comparisons, girls are the reference group (the first group); in pre–post comparisons, pre-course data is the baseline. 
If not indicated otherwise, all questionnaire items use a 5-point Likert scale and are converted into numeric values (1 = \emph{fully disagree}, 5 = \emph{fully agree}), with higher scores indicating stronger agreement.

\paragraph{RQ1: Perceived Knowledge\& Self-Efficacy} 
To assess whether students’ perceived knowledge regarding design thinking and software engineering improved during the course and differed by gender, we analysed the following items (\cref{tab:designthinkgindemos}): perceived experience with programming (\emph{KN01}) and software development (\emph{KN02}), familiarity with the overall software development process (\emph{KN03}) and design thinking process (\emph{KN04}). 
In addition, we measured self-efficacy (\emph{SE01}), i.e., students’ belief in their own ability to carry out software development tasks successfully.  
While perceived knowledge captures how much students think they know, self-efficacy reflects whether they feel capable of applying this knowledge in practice.

\paragraph{RQ2: Perceived Interest} 
To examine whether the course influenced students’ interest in software development and sustainability, we analysed six items. 
The questions \emph{IN01}--\emph{IN02} captured students’ general interest in software development and its relevance for their future profession, whereas \emph{IN03}--\emph{IN05} addressed students’ interest in sustainability, their participation in sustainability-related projects, and their perceived empowerment to contribute. 
In addition, \emph{IN06} was an open-text item, where responses were analysed using inductive thematic analysis to identify recurring ideas. This is especially well-suited for exploratory research since we do not have pre-defined categories. The qualitative dimension complemented the quantitative data by providing more insights into how students connect software with broader societal issues.

\paragraph{RQ3: Emotional Engagement} 
To capture students’ emotional engagement during the course, we asked them to rate eight emotions (e.g., happiness, boredom) for each design thinking phase presented. 
One author analysed the quantitative emotion ratings descriptively to identify overall distributions and gender differences. 
To better understand the reasoning behind these ratings, students also provided short justifications (\emph{EM02}). Although many comments were brief, we conducted inductive thematic analysis~\cite{braun2022}. 
This allowed us to group the reasons why certain phases were perceived as enjoyable or frustrating, thereby complementing the quantitative findings with explanatory insights. Selected quotes are reported to illustrate how specific course phases shaped students’ experiences. 

\paragraph{RQ4: Perceived Well-Being} 
To investigate whether participation in the ten-week course affected students’ broader well-being, we adopted six items from the Short Warwick–Edinburgh Mental Well-being Scale. These items (\emph{WB01}--\emph{WB06}) assessed optimism, perceived usefulness, relaxation, ability to deal with problems, clarity of thought, and closeness to others. Each item was measured on a 5-point Likert scale (1 = \emph{none of the time}, 5 = \emph{all of the time}), with higher scores indicating more frequent positive experiences.

\subsection{Positionality}
This study was conducted by an international research team consisting of two women and one man, all with European backgrounds. We have extensive experience in software engineering (education), including teaching at secondary, undergraduate, and graduate levels. Our prior work includes gender- and culture-sensitive teaching and sustainability-oriented education. We acknowledge that our expertise may have influenced interpretations.

\subsection{Threats to Validity}
\paragraph{Internal Validity.} 
One possible threat is the novelty effect: students may have reported positive emotions or learning outcomes simply because the design thinking approach and app prototyping were new to them. We mitigated this by spanning the course over ten weeks, which reduces short-term novelty effects, and by analysing reflections across all phases. Another threat is prior programming experience, which may have influenced engagement. We accounted for this by collecting baseline data on previous knowledge and by providing optional extensions during the design and programming phases. 
The presence of classroom teachers may have influenced responses, as well as the same female instructor leading all courses, particularly regarding gender stereotypes~\cite{sansone2017a}. We tried to mitigate this by ensuring survey anonymity and triangulating findings with qualitative reflections.

\paragraph{External Validity.} 
Our study was conducted in two schools within one country, which limits generalisability. Replication in other regions, cultural contexts, and age groups is needed to confirm the findings. Moreover, the course was facilitated by the same female instructor in both schools to ensure consistency. While this improves comparability, it may  limit transferability to other settings. Future work should investigate whether similar effects occur with different instructors, including regular classroom teachers.  
We also acknowledge the limited sample size. However, conducting a ten-week study in real classrooms is logistically challenging due to the tight school curriculum. Gaining access for such a long intervention is difficult, and therefore, the achieved scope and depth already provide meaningful insights.

\paragraph{Construct Validity.} 
We primarily relied on self-reported data from minors, which may be biased by social desirability or limited self-reflection~\cite{read2022c}. To mitigate this, we ensured anonymity and included qualitative reflection tasks to triangulate survey results. 
The longitudinal engagement over ten weeks, while a pedagogical strength, may have led to a drop-off in motivation or changes in classroom dynamics over time. We tracked session attendance and included a motivation check in each session reflection to monitor this.
Finally, cultural, socio-economic, and linguistic factors were not deeply captured in our data, limiting insights into intersectional experiences. Future research could include structured interviews or diary studies to address these dimensions.


\section{Results}
The following section presents findings from our mixed-methods evaluation, structured according to both research questions.

\subsection{RQ1: Knowledge and Self-Efficacy}

\begin{table}[t]
\centering
\small
\caption{Statistics of perceived knowledge and self-efficacy (RQ1, $p$-value, $\hat{A}_{12}$).}
\label{tab:experiences}
\begin{tabular}{lrrrr}
\toprule
& \multicolumn{2}{c}{Pre vs Post} & \multicolumn{2}{c}{Girls vs Boys} \\
Var. & Girls & Boys  & Pre & Post\\
\midrule
KN01 & \textbf{0.026, 0.375} & 0.957, 0.473 & 0.194, 0.403 & 0.714, 0.469 \\
KN02  & \textbf{0.014, 0.267} & 0.070, 0.340 & 0.377, 0.431 & 0.479, 0.553\\
KN03 & \textbf{0.001, 0.154} & \textbf{0.003, 0.171} & 0.720, 0.472 & 0.919, 0.491\\
KN04 & \textbf{0.005, 0.305} & 0.564, 0.478 & 0.264, 0.587 & 0.091, 0.622 \\ 
\midrule
SE01 & \textbf{0.046, 0.389} & \textbf{0.013, 0.233} & 0.165, 0.600 & 0.620, 0.459\\
\bottomrule
\end{tabular}
\end{table}

\begin{figure}
	\centering
	\includegraphics[width=1\linewidth]{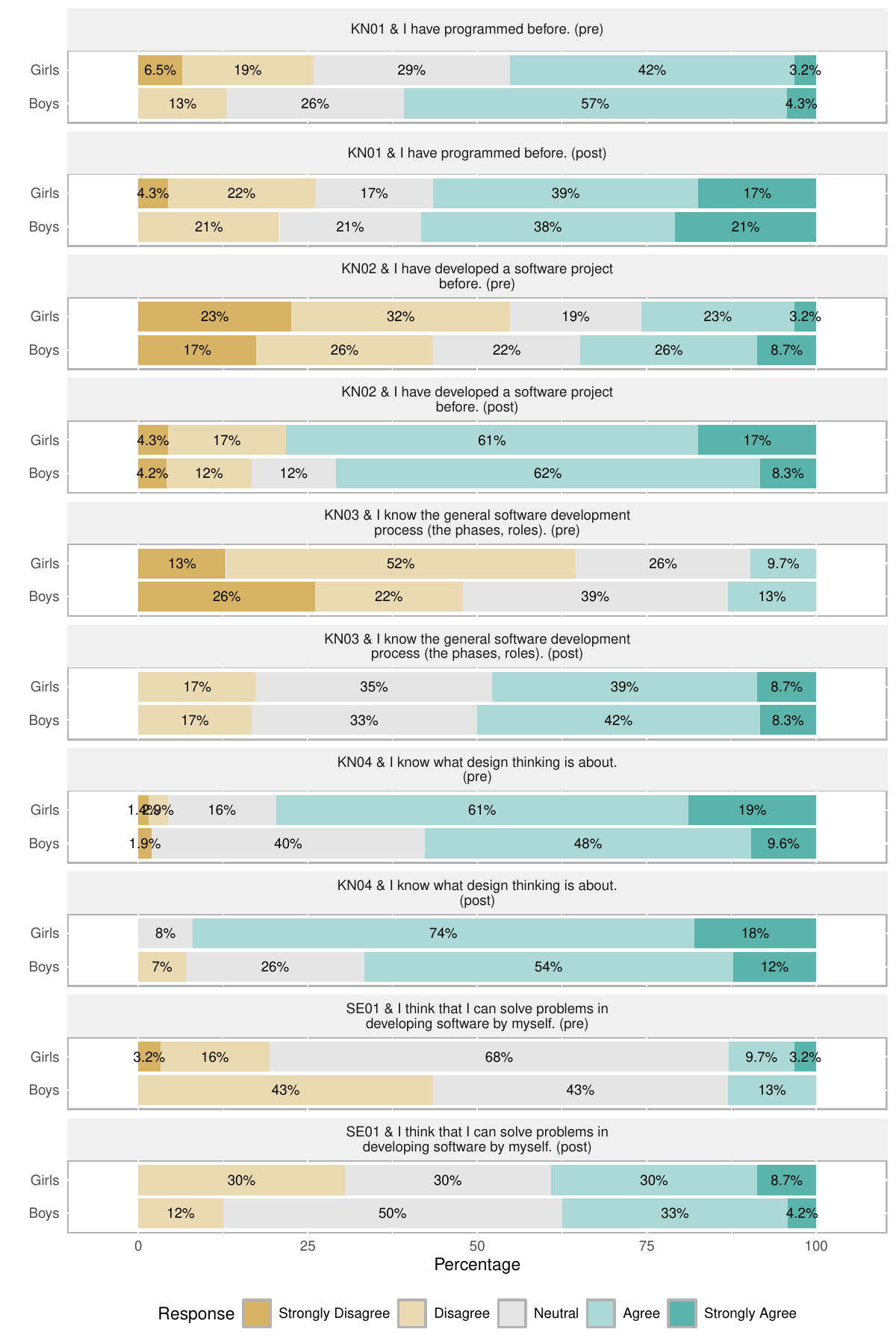}
	\caption{Distribution of perceived knowledge and self-efficacy.}
	\label{fig:experiences}
\end{figure}

\subsubsection{Perceived Knowledge (KN01--04)}
\Cref{fig:experiences} shows that boys entered the courses with more prior programming experience,\footnote{Teachers noted that none of the classes had prior programming instruction, so this experience (e.g., Scratch, Robot Karol) must be extracurricular.} while girls reported slightly stronger awareness of design thinking. Knowledge of software projects and processes was evenly distributed. \Cref{tab:experiences} shows no statistically significant differences.

Girls reported significantly improved experience in programming (\emph{KN01}, $p = 0.026, \hat{A}_{12} = 0.375$, \cref{tab:experiences}) and software project development (\emph{KN02}, $p = 0.014, \hat{A}_{12} = 0.267$), while boys showed a positive trend, although not statistically significant ($p = 0.957, \hat{A}_{12} = 0.473$; $p = 0.070, \hat{A}_{12} = 0.340$).
This may be due to greater prior programming experience, suggesting that the curriculum may  effectively support novices, especially girls who entered with less prior experience.

Both genders reported significant gains in their perceived understanding of the software development process (\emph{KN03}, girls: $p = 0.001, A12 = 0.154$; boys: $p = 0.003, \hat{A}_{12} = 0.171$, \cref{tab:experiences}).

For design thinking knowledge specifically, girls reported statistically significant improvement (\emph{KN04}, $p = 0.005, \hat{A}_{12} = 0.305$, \cref{tab:experiences}). The boys did not report significant changes (\emph{KN04}, $p = 0.564, \hat{A}_{12} = 0.478$). 
While boys maintained marginally higher knowledge levels post-course, the gap between genders narrowed substantially compared to pre-course differences (\cref{fig:experiences}). 

Overall, the design thinking approach seems to support foundational software knowledge~\cite{chatzinikolakis2014}, especially for learners with limited prior programming exposure, often girls, addressing traditional entry barriers in software engineering education. 

\subsubsection{Self-Efficacy (SE01)}
In contrast to prior literature~\cite{tellhed2022b}, boys reported lower initial self-efficacy despite their greater programming experience, possibly reflecting a more realistic awareness of challenges (\cref{fig:experiences}).

Both girls and boys reported significantly improved self-efficacy in solving software development tasks (\emph{SE01}, girls: $p = 0.046, \hat{A}_{12} = 0.389$; boys: $p = 0.013, \hat{A}_{12} = 0.233$, \cref{tab:experiences}).

\begin{tcolorbox}[colback=blue!10, colframe=blue!60]
\textbf{RQ1 Summary.} Girls perceived a knowledge increase across all areas, boys primarily in process understanding. Both improved self-efficacy and narrowed gender gaps.
\end{tcolorbox}

\begin{tcolorbox}[colback=blue!10, colframe=blue!60]
\textbf{RQ1 Lesson Learned.} Design thinking can boost students’ knowledge and confidence in software engineering.
\end{tcolorbox}

\subsection{RQ2: Perceived Interest in Sustainability and Software Development}
 \begin{table}[t]
\centering
\small
\caption{Statistics of interest (RQ2, $p$-value, $\hat{A}_{12}$).}
\label{tab:experiencesRQ2}
\begin{tabular}{lrrrr}
\toprule
& \multicolumn{2}{c}{Pre vs Post} & \multicolumn{2}{c}{Girls vs Boys} \\
Var. & Girls & Boys  & Pre & Post\\
\midrule
IN01 & \textbf{0.032, 0.289} & 0.133, 0.284 & 0.381, 0.566&  0.777, 0.475 \\ 
IN02 & \textbf{0.032, 0.293} & 0.058, 0.369 & 0.344, 0.426 & \textbf{0.030, 0.327} \\
IN03 & \textbf{0.020, 0.318} & 0.973, 0.527 & 0.371, 0.566&  0.166, 0.615 \\
IN04 & \textbf{0.022, 0.272} & 0.805, 0.458 & 0.243, 0.409&  0.227, 0.598 \\
IN05& \textbf{0.002, 0.270} & 0.172, 0.319 & 0.938, 0.507& 0.845, 0.517 \\
\bottomrule
\end{tabular}
\end{table}

\begin{figure}
	\centering
	\includegraphics[width=1\linewidth]{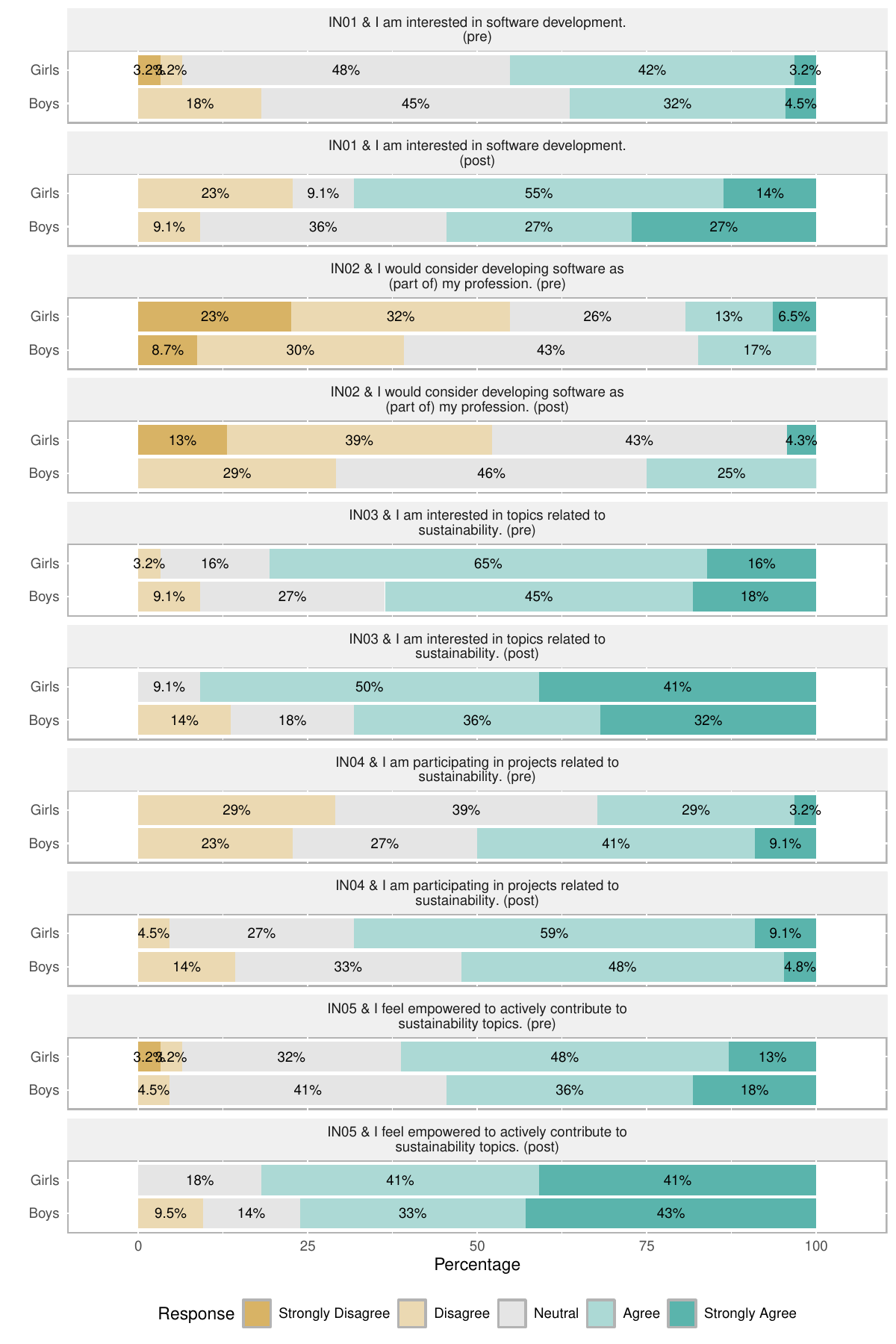}
	\caption{Distribution of interest in sustainability and software development.}
	\label{fig:interest}
\end{figure}

\subsubsection{Interest in Software Development (IN01--02)}
\Cref{fig:interest} shows clear differences between interest in boys and girls pre-course. Girls' interest in software development careers increased significantly (\emph{IN02}, $p = 0.032, \hat{A}_{12} = 0.293$, \cref{tab:experiencesRQ2}), while boys showed at least a positive trend ($p = 0.058, \hat{A}_{12} = 0.369$). 
A gender difference emerged post-course with boys considering a career more ($p = 0.030, \hat{A}_{12} = 0.327$, \cref{tab:experiencesRQ2}), suggesting that the intervention differentially impacted students’ career considerations.

\subsubsection{Interest in Sustainability (IN03--05)}
Girls also reported significant increases in interest in sustainability (\emph{IN03}, $p = 0.020, A12 = 0.318$, \cref{tab:experiencesRQ2}), participation in sustainability projects (\emph{IN04}, $p = 0.022, A12 = 0.272$), and empowerment to contribute to sustainability (\emph{IN05}, $p = 0.002, A12 = 0.27$), $p = 0.022, A12 = 0.272$). Boys showed no significant changes.

\subsubsection{Qualitative Data (IN06)}
The following key themes emerged from the thematic analysis of students' perceptions of the role of software engineering in sustainability. 

\textbf{Data-Driven Environmental Solutions.}  
Students initially saw software as supporting environmental data collection and management, e.g., one girl viewed software \emph{to provide management solutions to track how much waste we produce}~\textsuperscript{ID146}, or one boy regarded \emph{new software can help us manage our greenhouse gas emissions.}~\textsuperscript{ID70} 
Post-course, ideas became even more concrete with actual proposals as one boy, based on their project, suggest to \emph{build sensor networks that automatically collect environmental data and use machine learning to predict optimal energy usage patterns}~\textsuperscript{ID73}, with one girl noting software’s nature as \emph{software engineering can help with sustainability, because it's virtually done and doesn't require much resources to do. Therefore, it doesn't cause harm, and could potentially inspire sustainability in people.}~\textsuperscript{ID142} 

\textbf{Educational Applications.}  
Girls consistently emphasised education and awareness: \emph{Developing applications that educate people on sustainability and encourage them to participate in actions that aid people and the environment.}~\textsuperscript{ID145} Post-course, boys also incorporated educational projects, showing broader recognition of software’s teaching potential.

\textbf{Technological Innovation.}  
Initially, boys focused on general technologies, such as robots and AI, and girls on software solutions, such as trash-collecting systems at sea. Post-course, both genders integrated technical and social considerations equally into concrete solutions.

\textbf{Uncertainty $\rightarrow$ Concrete Ideas.}  
Pre-course uncertainty about the software’s role was common among girls and boys. Post-course, students proposed concrete, project-related ideas, often linked to their own or peers’ projects.

\textbf{Skepticism $\rightarrow$ Constructive Reflection.}  
Boys initially expressed direct skepticism (\emph{Na because they do only software}~\textsuperscript{ID45}), girls nuanced concerns (\emph{They can help promote sustainability, but their field is harder to make more eco-friendly.}~\textsuperscript{ID150}). 
Post-course, skepticism shifted to reflective thinking as one girl stated: \emph{Software alone isn't enough - you need to understand the real human problems first, like we learned, and then technology can be really powerful.}~\textsuperscript{ID150}

\textbf{Social Justice and Equity (Post-Course).}  
Students increasingly emphasised accessibility and equity, such as one girl explained: \emph{Create software that helps ensure solutions don't just benefit wealthy communities but are accessible to everyone}~\textsuperscript{ID154} and one boy mentioned \emph{Design platforms that strengthen voices from communities most affected by injustice and climate change.}~\textsuperscript{ID05}

\begin{tcolorbox}[colback=blue!10, colframe=blue!60]
\textbf{RQ2 Summary.} Girls’ interest in software development and sustainability increased significantly, while boys showed smaller gains.
\end{tcolorbox}

\begin{tcolorbox}[colback=blue!10, colframe=blue!60]
\textbf{RQ2 Lesson Learned.} Framing software engineering in social and environmental contexts engaged students, especially girls.
\end{tcolorbox}

\subsection{RQ3: Emotional Engagement}




\begin{figure}[t!]
	\centering
	\includegraphics[width=1\linewidth]{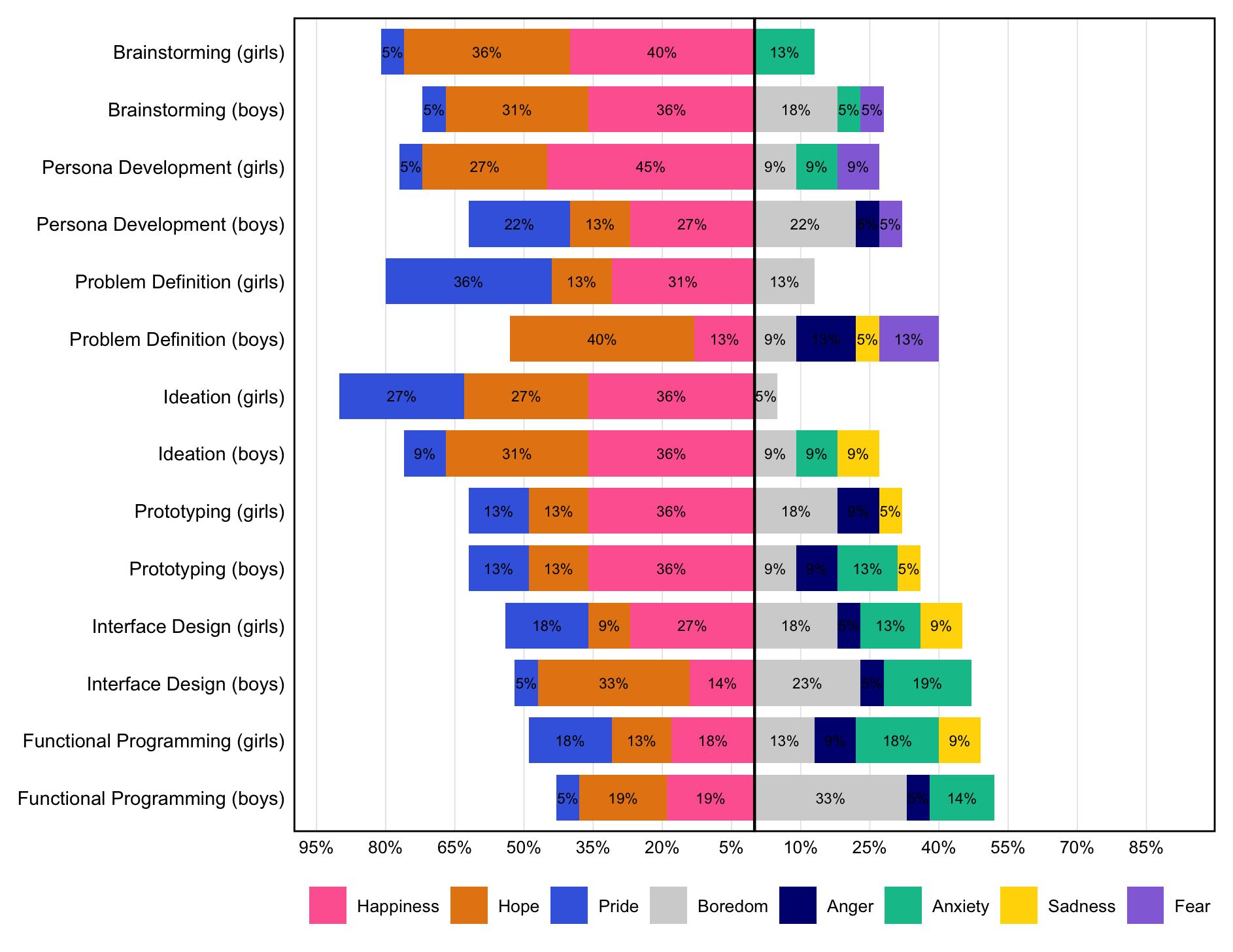}
	\caption{Reported emotions for the main course sessions.}
	\label{fig:emotionsphases}
\end{figure}

\subsubsection{Quantitative Results (EM01)}
While no significant gender differences emerged (all $p > 0.15$, $\hat{A}_{12}$ range: 0.222–-0.537), girls expressed more positive emotions overall, whereas boys reported boredom more frequently (\emph{EM01}, \cref{fig:emotionsphases}).

Quantitative ratings showed high levels of happiness and hope during the early phases (brainstorming, persona development, ideation). 
Pride primarily emerged during problem definition and prototyping. 
Engagement declined during later technical phases, particularly programming, with girls showing increased anxiety (18\%) and boys high boredom (33\%). Boys consistently reported higher levels of boredom than girls, most notably during the programming session.

Overall, positive emotions were strongest in creative and collaborative phases, while technical tasks led to boredom and anxiety, with minor gender differences.

\subsubsection{Qualitative Results (EM02)}
To better understand these patterns, we conducted a thematic analysis of students’ written reflections. The following themes emerged.

\textbf{Enjoyment of Creativity and Collaboration.} Early phases fostered happiness and hope through creative freedom and idea exchange. Students valued brainstorming and persona development for their openness and teamwork, as \emph{creating a person was fun.}~\textsuperscript{ID523} As one girl noted, \emph{I enjoyed hearing other perspectives, particularly in the brainstorming phase.}~\textsuperscript{ID460} Another described persona building as her \emph{favourite phase because [she] got to be creative.}~\textsuperscript{ID461} 

\textbf{Collaboration and Social Interaction.} Group work enhanced positive emotions and mitigated anxiety. Students highlighted pride in shared accomplishments: \emph{It felt good because we defined the problem together and I felt proud of our work}~\textsuperscript{ID478}, as well as valued discussion: \emph{I love to think and talk.}~\textsuperscript{ID482}

\textbf{Perceived Learning and Growth.} Many students associated pride and happiness with moments of visible progress or acquiring new skills. For example, a girl explained: \emph{Coding is where I learned the most because I was actually working with something new.}~\textsuperscript{ID461} Boys also emphasised overcoming initial difficulties: \emph{At the start I was so confused, but then I started to get it.}~\textsuperscript{ID505} One girl commented on interface design: \emph{I had never designed an app before, so I learned the most during this phase.}~\textsuperscript{ID460} 

\textbf{Repetition and Lack of Challenge.} Reports of boredom were concentrated in later phases, particularly among boys. Some students described phases as repetitive or not offering sufficient challenge: \emph{Designing the interface was boring because it felt repetitive}~\textsuperscript{ID523} as well as \emph{It got boring once I understood it.}~\textsuperscript{ID490}
However, many students also remarked that the programming was fun, since \emph{I really enjoy the logic of coding, so that was most enjoyable.}~\textsuperscript{ID460, girl} This suggests that while novelty fostered learning for some, others lacked adequate stimulation despite having complete freedom and being able to create additional or more complex features to the app.

\begin{tcolorbox}[colback=blue!10, colframe=blue!60]
\textbf{RQ3 Summary.} Positive emotions were strongest in creative, collaborative phases, while technical tasks caused boredom and anxiety, with minor gender differences.
\end{tcolorbox}

\begin{tcolorbox}[colback=blue!10, colframe=blue!60]
\textbf{RQ3 Lesson Learned.} Educators can sustain emotional engagement by combining creative and social activities with technical tasks rather than separate activities.
\end{tcolorbox}

\subsection{RQ4: Perceived Well-Being}
\begin{figure}
	\centering
	\includegraphics[width=1\linewidth]{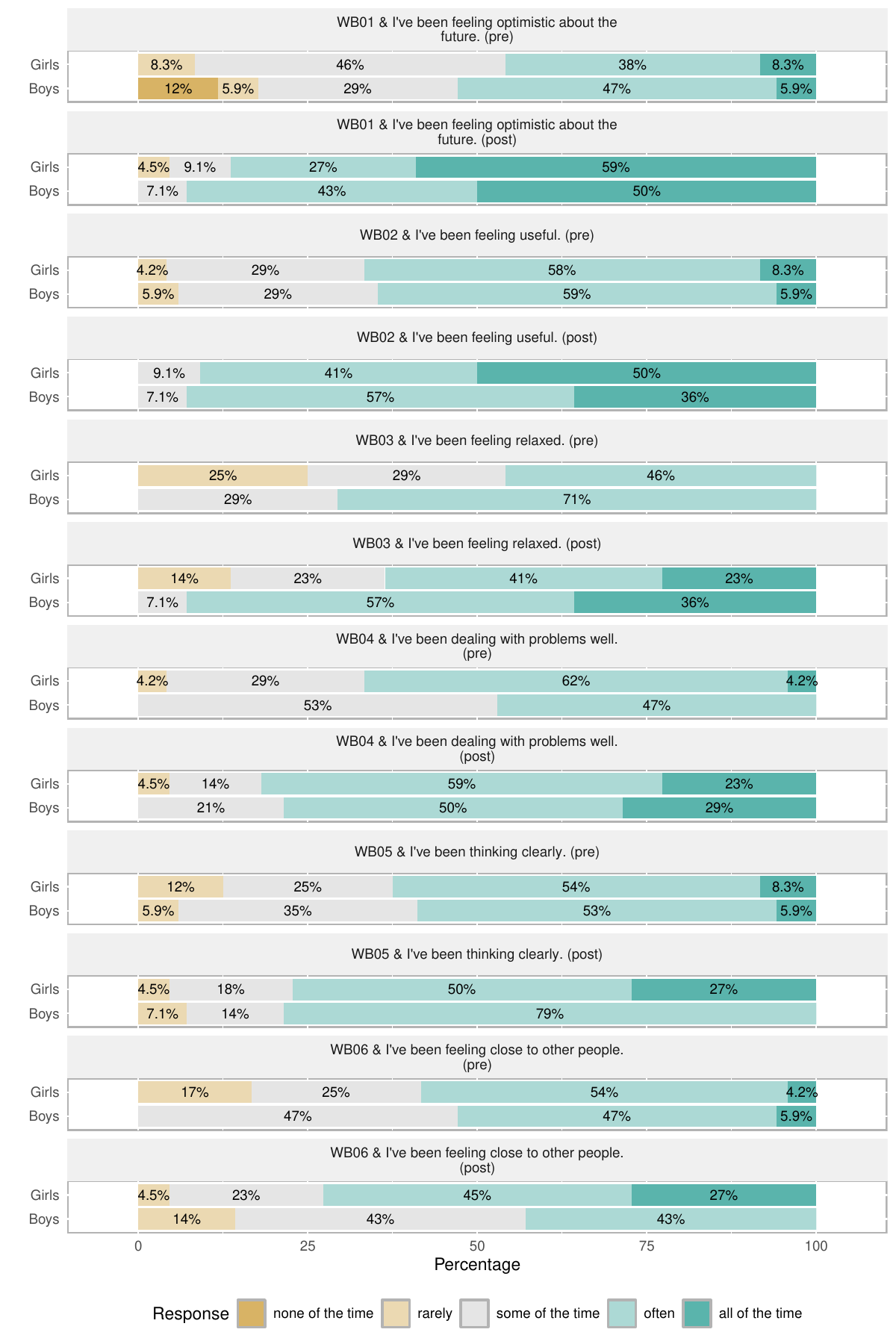}
	\caption{Distribution of perceived well-being.}
	\label{fig:wellbeingdistribution}
\end{figure}
\begin{table}[t]
\centering
\small
\caption{Statistics of well-being measures (RQ4, $p$-value, $\hat{A}_{12}$).}
\label{tab:wellbeing}
\begin{tabular}{lrrrr}
\toprule
& \multicolumn{2}{c}{Pre vs Post} & \multicolumn{2}{c}{Girls vs Boys} \\
Var. & Girls & Boys  & Pre & Post\\
\midrule
WB01 & \textbf{0.002, 0.141} & 0.078, 0.235 & 0.955, 0.506 & 0.798, 0.524 \\
WB02 & \textbf{0.007, 0.201} & 0.312, 0.340 & 0.822, 0.520 & 0.516, 0.560 \\
WB03 & \textbf{0.006, 0.227} & 0.062, 0.260 & 0.054, 0.339 & 0.091, 0.341 \\
WB04 & \textbf{0.034, 0.348} & \textbf{0.031, 0.190} & 0.238, 0.597 & 0.871, 0.484 \\
WB05 & \textbf{0.013, 0.289} & 0.625, 0.395 & 0.919, 0.510 & 0.242, 0.604 \\
WB06 & \textbf{0.046, 0.342} & 1.000, 0.500 & 0.838, 0.482 & \textbf{0.023, 0.714} \\
\bottomrule
\end{tabular}
\end{table}

\Cref{fig:wellbeingdistribution} again shows clear differences between
genders on well-being, and \cref{tab:wellbeing} reveals significant
improvements for girls across multiple dimensions following the design
thinking course.
In particular, girls showed statistically significant increases in optimism about the future (\emph{WB01}, $p = 0.002$, $\hat{A}_{12} = 0.141$), feelings of usefulness (\emph{WB02}, $p = 0.007$, $\hat{A}_{12} = 0.201$), relaxation (\emph{WB03}, $p = 0.006$, $\hat{A}_{12} = 0.227$), dealing with problems (\emph{WB04}, $p = 0.034$, $\hat{A}_{12} = 0.348$), clear thinking (\emph{WB05}, $p = 0.013$, $\hat{A}_{12} = 0.289$), and closeness to others (\emph{WB06}, $p = 0.046$, $\hat{A}_{12} = 0.342$). 

Boys demonstrated more limited improvements, with significant gains only in relaxation (\emph{WB03}, $p = 0.062$, $\hat{A}_{12} = 0.260$) and dealing with problems (\emph{WB04}, $p = 0.031$, $\hat{A}_{12} = 0.190$). 

Gender comparisons showed minimal differences before the start of the course, but post-course analysis revealed that girls reported significantly higher levels of feeling close to others compared to boys (\emph{WB06}, $p = 0.023$, $\hat{A}_{12} = 0.714$). 

\begin{tcolorbox}[colback=blue!10, colframe=blue!60]
\textbf{RQ4 Summary.} Girls showed overall gains in perceived well-being, while boys improved in relaxation and problem handling. Post-course, girls felt closer to others than boys. 
\end{tcolorbox}

\begin{tcolorbox}[colback=blue!10, colframe=blue!60]
\textbf{RQ4 Lesson Learned.} Design thinking can boost students’ well-being, especially girls, by combining computing tasks with collaboration and socially meaningful projects.
\end{tcolorbox}

\section{Discussion}
Our findings have implications for how we design and teach introductory software engineering courses.

\subsection{Empowering Girls}
Across all research questions, a consistent pattern emerges: design thinking makes software engineering feel meaningful and empowering. Girls benefited the most, showing significant gains in knowledge, self-efficacy, and positive emotions. Boys also improved, but more modestly and variably. This suggests that design thinking reduces barriers, connected potentially to negative stereotypes, that disproportionately affect girls, such as lack of belonging, low perceived relevance, and anxiety about technical skills.  

\textbf{Implication.} Educators can use human-centred, socially relevant approaches to make software engineering more inclusive.

\subsection{Self-efficacy, Mastery, and Culture}
The results of RQ1 showed that design thinking fosters strong mastery experiences~\cite{bandura1977a}, the most potent source of self-efficacy. 
Interestingly, boys with prior programming experience began with lower self-efficacy, perhaps reflecting how informal learning can erode rather than build confidence when many first encounter programming at home, which then can be frustrating.

Our culturally diverse group may explain the effects for girls, consistent with studies from non-Western contexts, in which gender gaps in confidence and interest in programming and software engineering careers are smaller~\cite{chen2024exploring,tellhed2022b}. 

\textbf{Implication.} Structured mastery experiences can build self-efficacy for all students, but may be culturally sensitive.

\subsection{Emotions and Gendered Engagement}
These improvements from RQ1 also correspond closely with RQ3’s findings on emotional engagement: positive emotions such as happiness and hope peaked during the early, creative stages of the course, where persona building and prototyping were described as fun. This suggests that emotional engagement is not merely a by-product but a foundational driver of learning and motivation. 

In contrast, the technical implementation triggered anxiety in girls and boredom in boys, mirroring prior findings~\cite{zhang2022b} that collaborative prototyping fosters enjoyment, whereas solitary programming reduces engagement. 
For boys, two factors likely contributed to boredom: (1) greater prior exposure to block-based tools from extracurricular activities, making App Inventor feel repetitive, and (2) teamwork imbalances, where one student programmed while others watched, despite being tasked to do research, such as design decisions or looking for bugs. 
For girls, a limited programming background might have made the implementation phase more challenging and anxiety-provoking. This emotional downturn may threaten sustained engagement, even though students continued to report learning gains. As one girl reflected: \emph{I think coding I learned the most because I was actually working with something new... My favourite phase was the persona building because I got to be creative.}~\textsuperscript{ID461} 

These patterns reveal differentiated needs. Girls thrived in creative, collaborative phases, while boys required novelty and technical challenge to stay engaged. As one boy put it: \emph{Loved having fun in the coding blocks. At the start, tho I was so confused, but then I started to get it.}~\textsuperscript{ID505} 
This suggests that differentiated instruction that dynamically adjusts to prior knowledge and interests may improve outcomes for all genders. Educators must consider  what students learn and how they feel while learning. One girl captured this connection: 
\emph{The app is going really well!! I figured out a few things that really had me stumped, but now I am super happy!!}~\textsuperscript{ID399}

\textbf{Implication:} Educators should design programming tasks that sustain positive emotions by balancing creativity, (technical) novelty, \emph{and} collaboration.

\section{Conclusions and Future Work}
This study provides empirical evidence that integrating design thinking into software engineering education engages and empowers secondary school students and supports gender equity. Human-centred approaches foster significant gains in girls’ self-efficacy, interest, and well-being while also benefiting boys. The distinct gendered preferences observed highlight the importance of balancing creative and technical activities to broaden participation.
Future work could explore technical variations with tool choices while maintaining design thinking's creative framework, integrating it into traditional programming education, like text-based programming, and systematically examining gender and other diversity dimensions to support
equitable computing education.

\bibliographystyle{ACM-Reference-Format}
\bibliography{references}

\end{document}